\documentclass[twocolumn]{aastex631}

\usepackage{graphicx}

\begin{document}

\title{The Tip of Red Giant Branch Distances to Nearby Dwarf Galaxies WLM and Sextans A with JWST}

\author{Ziming Yan}
\affiliation{CAS Key Laboratory of Optical Astronomy, National Astronomical Observatories, Chinese Academy of Sciences, Beijing 100101, People's Republic of China}
\affiliation{School of Astronomy and Space Science, University of the Chinese Academy of Sciences, Beijing 100049, People's Republic of China}

\author[0000-0003-4489-9794]{Shu Wang}
\affiliation{CAS Key Laboratory of Optical Astronomy, National Astronomical Observatories, Chinese Academy of Sciences, Beijing 100101, People's Republic of China}

\author[0000-0001-7084-0484]{Xiaodian Chen}
\affiliation{CAS Key Laboratory of Optical Astronomy, National Astronomical Observatories, Chinese Academy of Sciences, Beijing 100101, People's Republic of China}
\affiliation{School of Astronomy and Space Science, University of the Chinese Academy of Sciences, Beijing 100049, People's Republic of China}
\affiliation{Institute for Frontiers in Astronomy and Astrophysics, Beijing Normal University, Beijing 102206, People's Republic of China}

\author{Licai Deng}
\affiliation{CAS Key Laboratory of Optical Astronomy, National Astronomical Observatories, Chinese Academy of Sciences, Beijing 100101, People's Republic of China}
\affiliation{School of Astronomy and Space Science, University of the Chinese Academy of Sciences, Beijing 100049, People's Republic of China}

\correspondingauthor{Shu Wang}
\email{shuwang@nao.cas.cn}

\begin{abstract}
Distance measurements to extragalactic systems that are both accurate and precise are cornerstones of modern astrophysics, underpinning the calibration of standard candles and the determination of the Hubble constant. Dwarf galaxies, such as Wolf--Lundmark--Melotte (WLM) and Sextans A, provide valuable laboratories for testing distance scales across different stellar populations. In this work, we utilize the high sensitivity and spatial resolution of the James Webb Space Telescope (JWST) to measure the distances to WLM and Sextans A using the tip of the red giant branch (TRGB) method. Adopting the TRGB absolute magnitude calibrated by NGC 4258, we determine distance moduli of $\mu_{\mathrm{0,WLM}} = 24.977 \pm 0.018 (\mathrm{stat}) \pm 0.056 (\mathrm{sys})$ mag for WLM and $\mu_{\mathrm{0,SexA}} = 25.740 \pm 0.011 (\mathrm{stat}) \pm 0.057 (\mathrm{sys})$ mag for Sextans A. Our results are consistent within a 3\% distance uncertainty with previous measurements based on TRGB, Cepheids, and J-Region Asymptotic Giant Branch (JAGB) methods. With improved distance measurements in the future, these two galaxies have the potential to serve as additional anchor points for TRGB calibration, aiming to reduce the TRGB-based distance uncertainty to below 2\%.
\end{abstract}


\section{Introduction} \label{sec:intro}

Dwarf galaxies are important members of the Local Group and probably play a significant role in the formation of large galaxies \citep{1978MNRAS.183..341W}. They usually have simple star formation histories \citep{2014ApJ...789..147W}, low and spatially homogeneous metallicities \citep{2009ARA&A..47..371T}, making them suitable for the study of stellar populations. Due to their proximity and well-resolved characteristics, more accurate distances can be measured with less influence of crowding effects. Many old dwarf galaxies also have star-forming regions and different types of standard candles like Cepheids, RR Lyraes and the tip of red giant branch (TRGB), which is beneficial for cross-checking the distance measurements.

TRGB has become a reliable method for measuring the distances of nearby galaxies \citep{2019ApJ...882...34F,2021ApJ...919...16F}. It marks the helium flash in old low-mass stars. The stars keep burning hydrogen into helium as they evolve along the red giant branch (RGB). At the end of the RGB phase, the temperature of the degenerate helium core reaches the critical point of helium ignition and thus the helium flash begins. The stars then evolve to the core helium-burning horizontal branch, and the decrease in their luminosity results in a sharp cut-off of the RGB luminosity function, or the so-called TRGB. The luminosity of TRGB depends weakly on composition or stellar mass, and is nearly constant in specific filters, especially the \textit{I} band \citep{1990AJ....100..162D,1993ApJ...417..553L}, which makes it a standard candle comparable to Cepheids or RR Lyraes. The theoretical frame of TRGB is already clear \citep{1997MNRAS.289..406S,2017A&A...606A..33S} and it has been applied to measure distances to many nearby galaxies \citep{2009AJ....138..332J,2017ApJ...834...78M,2006AJ....131.1361K,2013ApJ...773...13L}.

The James Webb Space Telescope (JWST) \citep{2023PASP..135d8001R,2023PASP..135b8001R} has greater sensitivity and higher spatial resolution in the near- and mid-infrared, compared to the Hubble Space Telescope (HST). These can help resolve the distant crowded fields located tens of Mpc away, where the crowding effect is becoming significant for HST, and lower the photometric errors. In the $I$ band, the TRGB magnitude remains relatively constant for low-metallicity/blue RGB stars. Beyond about 1 micron there is a measurable brightening of the TRGB magnitudes with increased colors, which can be calibrated (e.g., \citet{2025arXiv250311769H}). Moreover, the extinction in the infrared bands is significantly smaller than that in the optical bands, which can reduces errors in the distance determinations. HST has exploited the TRGB method extensively using the optical $F814W$ filter (similar to $I$-band) to measure distances out to $\sim$20 Mpc. With JWST it is likely that the TRGB distance scale can be extended out to $\sim$50 Mpc.

Recent measurements of the Hubble constant using different TRGB host galaxies show significant discrepancies \citep{2024arXiv240806153F,2025arXiv250311769H,2025arXiv250408921L}, indicating that the TRGB magnitude may still be affected by photometric quality and stellar population properties (however, in support of past application, see the extensive numerical simulations by \citet{2023AJ....166....2M}), and thus suggests further investigation.

Dwarf galaxies like Wolf–Lundmark–Melotte (WLM) and Sextans A are among the smallest and most metal-poor star-forming systems in the local Universe. WLM is an archetypal isolated dwarf in the outskirts of the Local Group at a distance of about 0.97 Mpc \citep{2025ApJ...981..153C}. It lies $\sim$800 kpc from both the Milky Way and M31, well outside their virial radii, and is thought to be on its first infall into the Local Group. Sextans A, similarly, is a gas-rich dwarf irregular galaxy located roughly 1.3 Mpc from us \citep{2003AJ....125.1261D}, likely on the periphery of the Local Group or in the nearby NGC 3109 association. Both WLM and Sextans A have extremely low heavy-element abundances (on the order of $0.1Z_{\odot}$ or less), and they host mixed stellar populations ranging from ancient, metal-poor red giants to young, massive stars, allowing inter comparison of the various distance indicators.

In this paper, we use JWST data to measure the TRGB magnitudes of WLM and Sextans A and compare the resulting distances with those derived from other methods. This paper is organized as follows. Section 2 describes the JWST observations of WLM and Sextans A. Section 3 introduces the data reduction and photometry procedures. In Section 4, we present the color–magnitude diagrams and the determination of the TRGB magnitude in each galaxy. Section 5 includes the resulting distance determinations, comparisons with previous distance estimates and discussion. Section 6 provides a summary of this work.

\section{Data} \label{sec:data}
The data used in this work is from the JWST Cycle 1 Early Release Science (ERS) Program 1334 (PI: Daniel R. Weisz) for WLM and General Observer (GO) Program 1619 (PI: Martha L. Boyer) for Sextans A. The split fields of the JWST NIRCam enable us to simultaneously observe the star-forming disks as well as the halo regions of our targets. All the {\it JWST} data used in this work can be found in MAST: \dataset[10.17909/a15x-y010]{http://dx.doi.org/10.17909/a15x-y010}.

The data for WLM includes only one NIRCam observation with the F090W, F150W, F277W and F444W filters while the data for Sextans A includes two NIRCam observations, one using the FULL array and one using just Module B, both taken with 4 filter pairs. However, we only use the F090W and F150W short wavelength (SW) images in this study considering the higher resolution of SW filters and the consistency of the TRGB magnitude across different colors in the F090W band.

\begin{figure*}[]
\plotone{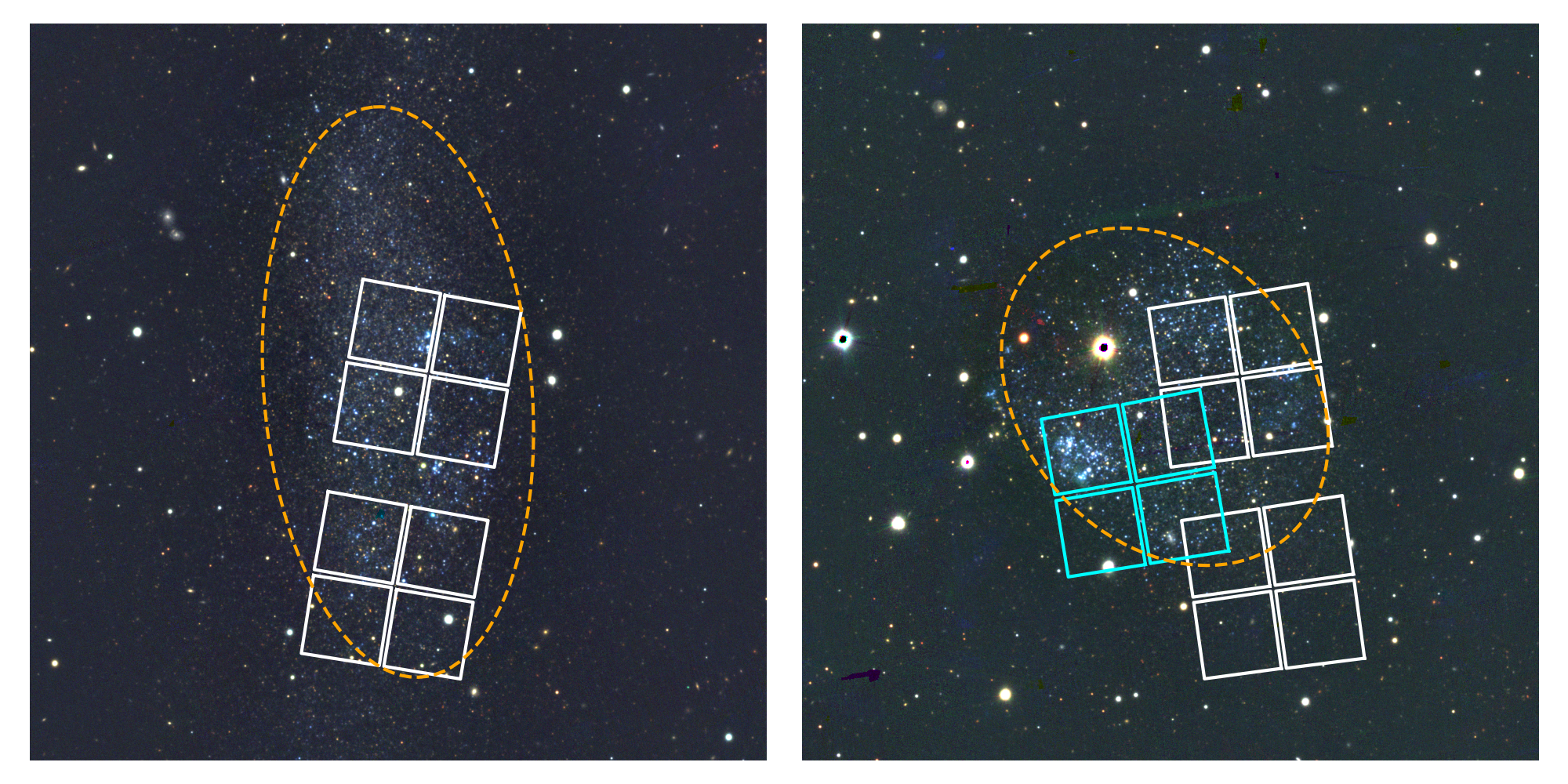}
\caption{Footprints of our NIRCam observations of WLM (Left) and Sextans A (right), overlaid on 10$^\prime \times$10$^\prime$ background images from the Pan-STARRS1 survey. The white and cyan boxes in the right panel represent the footprints of the observations ``o014\_t041" and ``o015\_t040" respectively. The 25th \textit{B}-band magnitude isophotes are shown as the dashed orange lines. \label{fig:footprint}}
\end{figure*}

The data for WLM and Sextans A were reduced with the jwst\_1303.pmap and jwst\_1293.pmap CRDS context files respectively, with no significant changes related to our NIRCam imaging compared to the versions used in \citet{2024ApJ...966...89A}. Minor differences between CRDS versions have negligible impact on the photometric measurements critical for TRGB determination. Detailed information about the observations is provided in Table \ref{tab:obslog} and the footprints are shown in Figure \ref{fig:footprint}, overlaid on 10$^\prime \times$10$^\prime$ background images from the Pan-STARRS1 survey.


\begin{table*}[]
\centering
\caption{Observation log for the data used in this work \label{tab:obslog}}

\begin{tabular}{cccccc}
\hline
\hline
Date & Target & Program ID & Observation ID & Filter & Exp. Time [s] \\
\hline
2022-07-23,24 & WLM & 1334 & o005\_t002 & F090W & 7623×4 \\
2022-07-24 & WLM & 1334 & o005\_t002 & F150W & 5927×4 \\
2023-04-15 & Sextans A & 1619 & o014\_t041 & F090W & 311×4 \\
2023-04-15 & Sextans A & 1619 & o014\_t041 & F150W & 311×4 \\
2023-04-15 & Sextans A & 1619 & o015\_t040 & F090W & 311×4 \\
2023-04-15 & Sextans A & 1619 & o015\_t040 & F150W & 311×4 \\ 
\hline
\end{tabular}

\end{table*}

\section{Data Reduction} \label{sec:data-reduction}

To obtain high-quality and robust star catalogs, we performed point spread function (PSF) photometry on the raw images calibrated with the JWST pipeline. We used the software DOLPHOT \citep{2000PASP..112.1383D,2016ascl.soft08013D}, a widely used crowded-field stellar photometry package, along with its JWST/NIRCam module \citep{2024ApJS..271...47W}. We perform PSF photometry on Stage 2 F090W and F150W images (*cal.fits) simultaneously for each observation, with the corresponding Stage 3 F150W image (*i2d.fits) as the reference frame.

The versions of DOLPHOT (December 1, 2024 release) and NIRCam PSFs (February 4, 2024 release) we used are both up-to-date. In the latest release of NIRCam PSFs, higher PSF spatial sampling was included to better match empirical PSFs. We noticed that the DOLPHOT NIRCam module adopted the Sirius-Vega system as default zeropoints instead of the Vega-Vega system since October 11, 2023, which would bring a small difference of $\sim$0.025 mag in the F090W magnitude, and thus we transformed the zero points to the Vega-Vega system by substituting filters\_vega.dat for filters.dat in the DOLPHOT NIRCam module to keep consistent with \citet{2024ApJ...966...89A}. We used the "-etctime" flag in the pixel masking process, which set the DOLPHOT exposure time in line with JWST ETC values. The reduction parameters we took for our photometry were the same as recommended in \citet{2024ApJS..271...47W}.


The original photometric catalogs provided by DOLPHOT are unculled, containing sources of low quality. We cull the catalogs with the criteria based on \citet{2024ApJS..271...47W} by applying all of the following selection criteria simultaneously: (1) Crowding $\leq 0.5$; (2) $(\text{Sharpness})^2 \leq 0.01$; (3) S/N $\geq 5$; (4) Error Flag $\leq 2$; (5) Object Type $\leq 2$. The S/N limit adopted here is slightly different from that in \citet{2023RNAAS...7...23W}. However, it won't affect our TRGB measurement, because the TRGB magnitudes of our targets are at least 5 mag brighter than the magnitude limit in F090W, and within the magnitude range near TRGB, the sources detected all have exceptionally high signal-to-noise ratio. All the criteria were applied to both bands except the object type.

\begin{figure}[]
\includegraphics[width=\columnwidth]{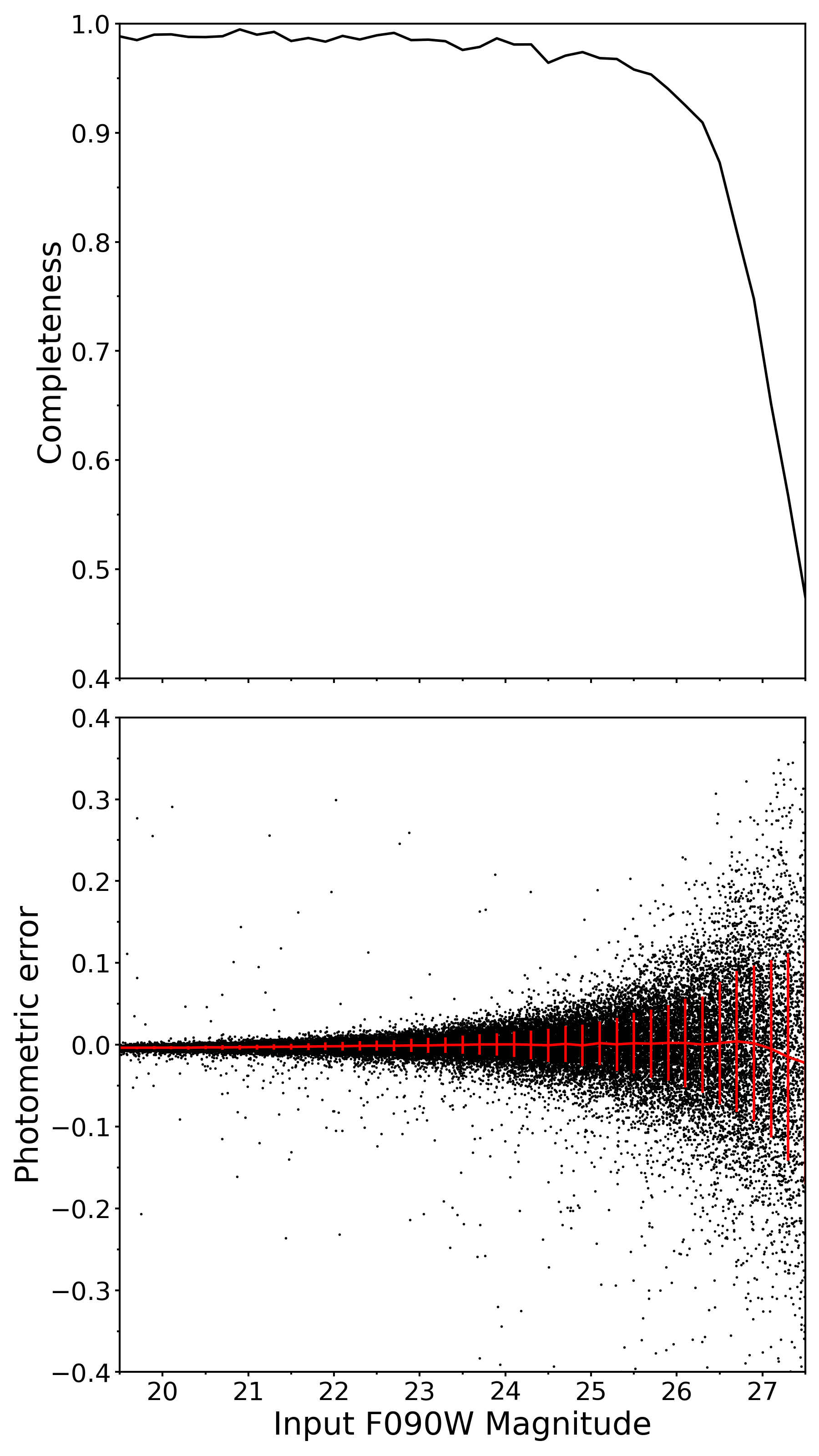}
\caption{Results of artificial star tests for Sex A. Top: Completeness curve as a function of input F090W magnitudes, defined as the ratio of the number of stars that were successfully detected and passed the quality criteria to the number of artificial stars injected. Bottom: The red line represents the photometric bias and error, with the background black dots representing the F090W magnitude differences between the output and input magnitude of each recovered star. \label{fig:photerror}}
\end{figure}

We also performed artificial star tests (ASTs) with DOLPHOT to evaluate the levels of completeness, photometric error and bias of our photometry. We injected 100000 artificial stars to the raw images for each chip, one at a time with a uniform spatial distribution, and repeated the photometric process to recover them. The same quality criteria were applied to the output catalogs of ASTs. The results of completeness and photometric error for Sextans A are shown in Figure \ref{fig:photerror}. As we can see from the figure, the completeness remains about 98.8\% until $\sim$3 magnitudes below TRGB, and the photometric error near the TRGB magnitude is only $\sim$0.003 mag. The results are similar in the case of WLM. For the 1.2\% of artificial stars that were not recovered, about 90\% are located at the edges of the detector chips, while the remaining losses are mainly due to artificial stars falling close to bright stars.

\section{TRGB Measurement} \label{sec:trgbmeasurement}

The method we used to measure the TRGB magnitude is the Sobel edge-detection algorithm, which was first introduced by \citet{1993ApJ...417..553L} and further developed by many works \citep{1995AJ....109.1645M,1996ApJ...461..713S,2008ApJ...689..721M}. Before applying the algorithm mentioned above, it is necessary to apply some restrictions to our data.

\subsection{Spatial Selection} \label{sec:spatial}


It is important to run TRGB measurements within the halo regions of galaxies due to greater crowding effects, larger extinction and higher risks of contamination by young- and intermediate-age populations in the disk regions. The cut was usually made by limiting the sample to the exterior of an ellipse or circle \citep{2006AJ....132.2729M,2017ApJ...845..146H,2018ApJ...858...11M}, or to the two outmost NIRCam SW chips that are the furthest from the galaxy center for simplicity \citep{2024ApJ...966...89A,2024arXiv240816810A}.

As shown in Figure \ref{fig:footprint}, the majority of our observational footprints falls on the star-forming disks of both galaxies, which was designed as the original goals of the two programs and not conducive to our TRGB measurements. We tried the 25th \textit{B}-band magnitude isophotes ($D_{25}$) from \citet{2011ApJS..197...21H} (see the blue dashed lines in Figure \ref{fig:footprint}) following \citet{2024ApJ...966...89A}. For WLM, however, this selection ended with the number of RGB stars inadequate for measuring TRGB. Instead we chose the two outermost SW chips (B1 and B2), which was acceptable considering the faintness and low extinction of the dwarf galaxy. For Sextans A, we kept the $D_{25}$ selection since there were two observations covering relatively a larger spatial area.


\begin{figure*}[]
\plotone{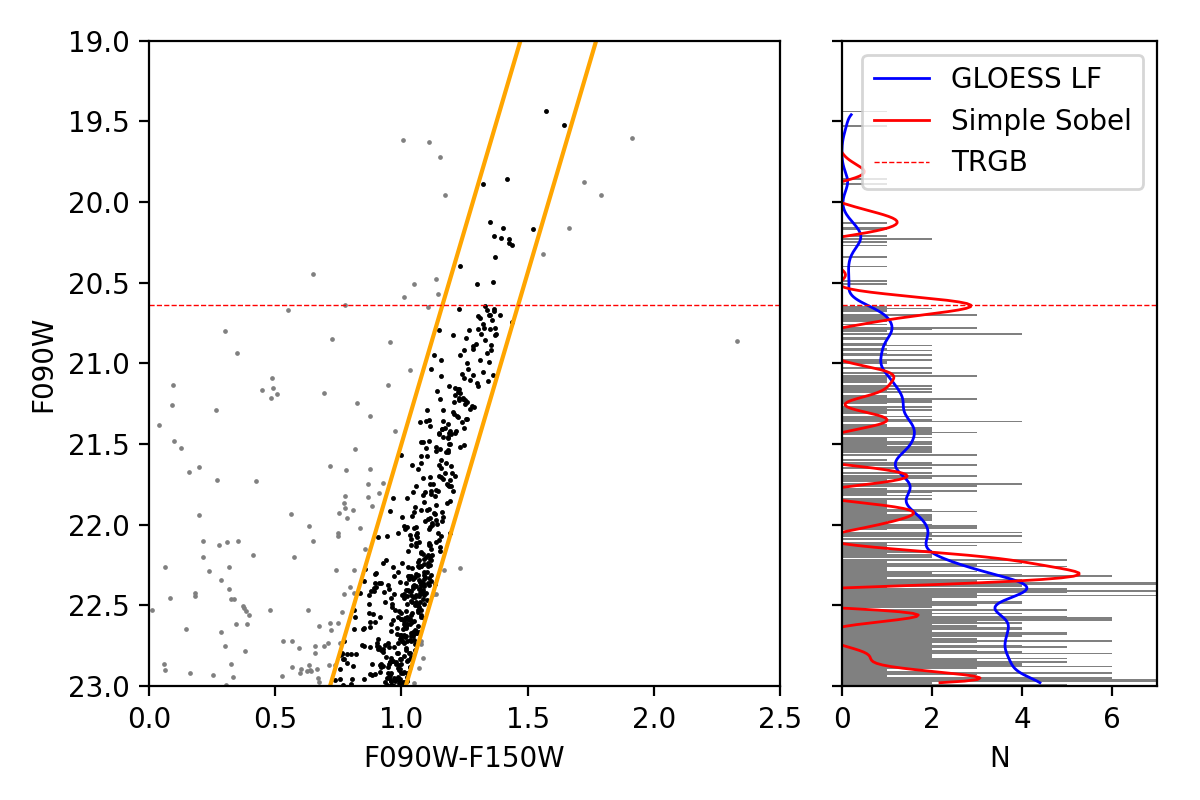}
\caption{Left: The color-magnitude diagram of the RGB region for WLM. The color band is plotted as the orange lines, while the black and grey dots represent stars that passed and failed the color selection, respectively. Right: The Sobel edge-detection results for WLM. The red solid line is the edge-detection response of the smoothed luminosity function (blue solid line). The first prominent peak of the Sobel response, or the location of TRGB, is designated by the red dashed horizontal lines in both panels. \label{fig:wlmtrgb}}
\end{figure*}

\subsection{Color Selection} \label{sec:color}

\begin{figure*}[]
\plotone{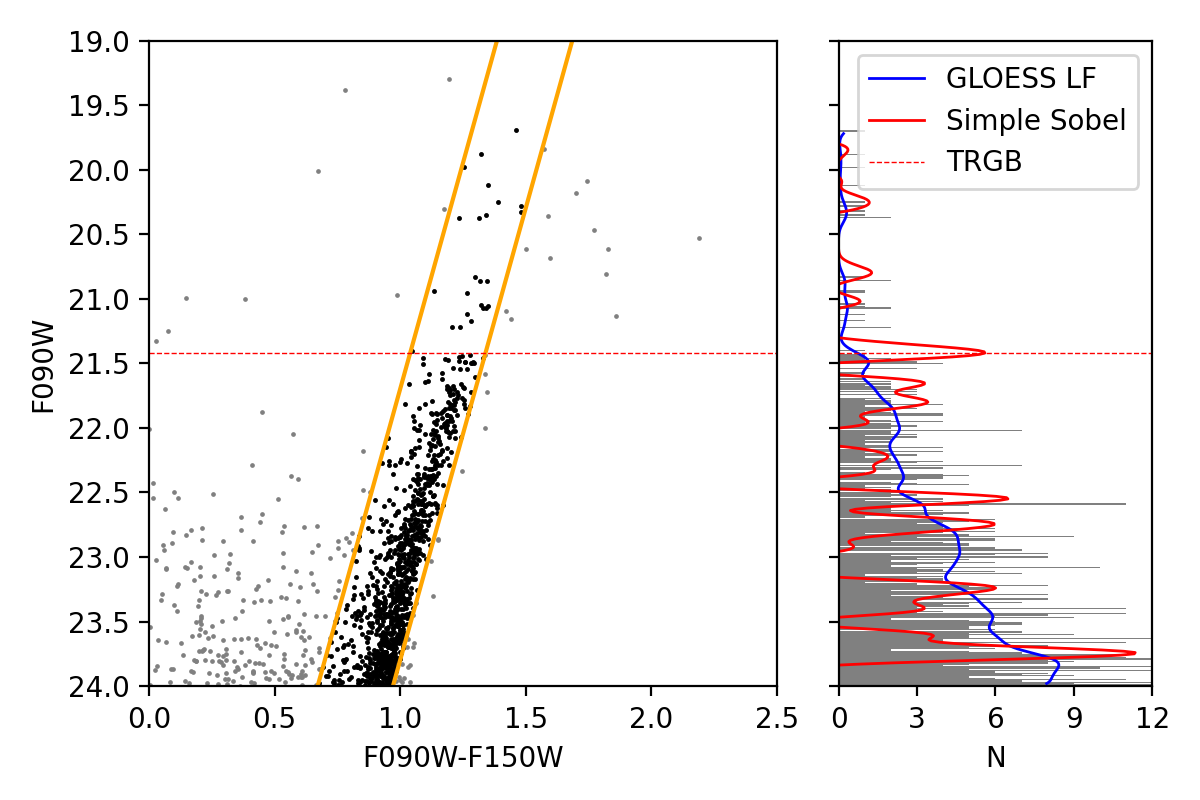}
\caption{Same as Figure \ref{fig:wlmtrgb}, for Sextans A. \label{fig:sextrgb}}
\end{figure*}

Color selection is usually applied to include only stars in a well defined RGB and reduce the influence from young giants that locate blueward of the RGB. \citet{2024ApJ...966...89A} presented that the TRGB magnitude in F090W varies little with metallicity ($< 0.02$ mag) over the color range of 1.15 $<$ (F090W-F150W) $<$ 1.75 mag for a 10 Gyr population and the variation with ages is almost negligible ($< 0.005$ mag). Due to the proximity and low metallicity of the two galaxies, the RGBs in their color-magnitude diagrams appear very narrow and clear. Thus we adopted a color band selection with a band width of 0.3 mag in each color-magnitude diagram covering the most RGB stars (see the left panels in Figure \ref{fig:wlmtrgb} and \ref{fig:sextrgb} and the orange lines therein). The color ranges at the corresponding TRGB magnitudes coincide with the recommended values, except a little excess on the blue end for Sextans A which hardly affects our results owing to the lack of young populations in our fields.


\subsection{Sobel Edge Detection} \label{sec:sobel}


We conducted the Sobel edge-detection procedure to the stars which passed the spatial and color selections. The color-magnitude diagrams of the RGB regions and the results of Sobel edge-detection are shown in Figure \ref{fig:wlmtrgb} and \ref{fig:sextrgb}. Initially, we binned the sample by F090W magnitude to obtain a F090W luminosity function with a bin width of 0.01 mag and smoothed the discrete luminosity function with  Gaussian locally-weighted regression smoothing (GLOESS) algorithm \citep{2004AJ....128.2239P}, plotted as the background gray histograms and the blue solid lines respectively in the right panels. The smoothing scale $\sigma_{s}$ we took was 0.06 mag. 
Then we applied the simple Sobel edge detection with a Sobel kernel of [-1,0,1], which was equivalent to computing the discrete first-order derivative of the luminosity function. The first significant peak of the Sobel edge-detection response was identified as the location of TRGB (see the red dashed horizontal lines in both figures).

To determine the means and uncertainties of TRGB magnitudes, we executed 1000 Monte Carlo trials with bootstrap resampling, adding Gaussian random errors based on the results of ASTs in each trial. In the final distribution of TRGB magnitudes, there were some peaks brighter than the true TRGB due to the noise caused by a few AGB stars, and simultaneously some peaks much fainter due to the relative sparseness of the RGB populations. Nonetheless, the peak at the estimated TRGB magnitude was clear and significant, which we fitted with a Gaussian after prior exclusion of obvious outliers. We took the corresponding mean and standard deviation of the Gaussian distribution as the mean magnitude and statistical uncertainty of TRGB. The final apparent TRGB magnitudes for our targets via the Sobel edge-detection algorithm were $m_{\mathrm{TRGB}} = 20.643 \pm 0.018$ mag for WLM and $21.415 \pm 0.011$ mag for Sextans A. We also examined the variations of TRGB magnitudes with the smoothing scale $\sigma_{s}$ and found the variations less than 0.01 mag for $\sigma_{s}$ ranging from 0.04 to 0.12, which proved the stability of our results.

\section{TRGB Distances and comparison with literature} \label{sec:distance}

To convert the apparent magnitudes to the distance moduli, we adopted the absolute magnitude of TRGB in F090W from \citet{2024ApJ...966...89A}, which was calibrated by NGC 4258, a water megamaser host galaxy with a precise geometric distance \citep{2013ApJ...775...13H,2019ApJ...886L..27R}. Different values were recommended depending on the specific measurement methodologies and here we took $M^{\mathrm{F090W}}_{\mathrm{TRGB}} = -4.377 \pm 0.033 (\mathrm{stat}) \pm 0.045 (\mathrm{sys})$ mag. The F090W band extinction coefficient, $A_{\mathrm{F090W}}/E(B-V) = 1.4021$, was calculated based on the adjusted reddening law with $R_V = 3.1$ \citep{2019ApJ...877..116W,2024ApJ...964L...3W}. With $E(B-V) = 0.0308$ mag (WLM) and 0.0374 mag (Sextans A) from \citet{2011ApJ...737..103S}, the extinction values were $A_{\mathrm{F090W}} = 0.0432$ mag and $0.0524$ mag, respectively. We considered 20\% error in the extinction determination. Thus we determined the distance moduli $\mu_{\mathrm{0,WLM}} = 24.977 \pm 0.018 (\mathrm{stat}) \pm 0.056 (\mathrm{sys})$ mag and $\mu_{\mathrm{0,SexA}} = 25.740 \pm 0.011 (\mathrm{stat}) \pm 0.057 (\mathrm{sys})$ mag.

\begin{figure}[]
\centering
\includegraphics[width=\linewidth]{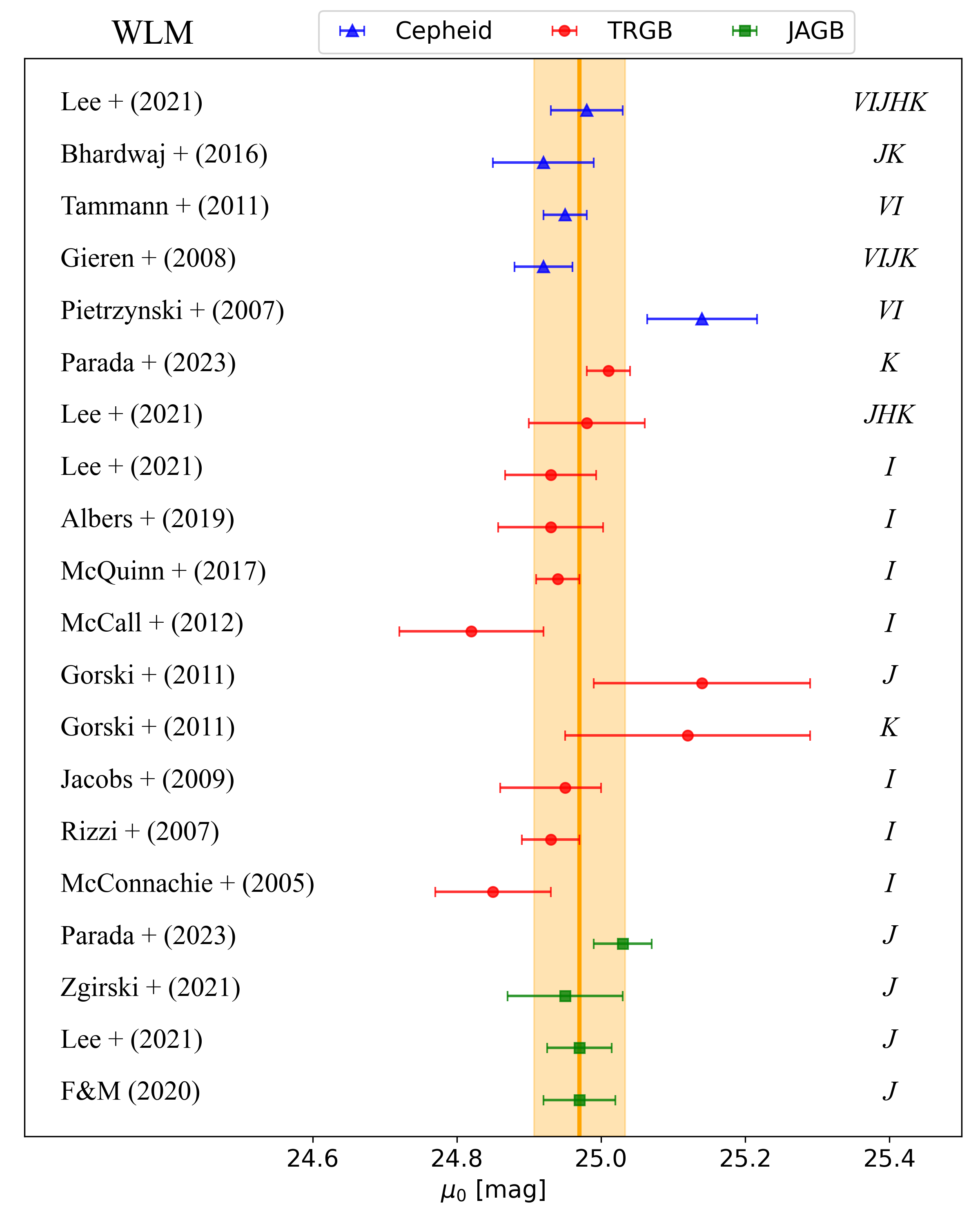} 
\caption{Distances for WLM in this work and literature. The orange vertical line indicates the mean of the  distance modulus from this work, with the shaded region representing the 1$\sigma$ error range. References: \citet{2005MNRAS.356..979M}, \citet{2007AJ....134..594P}, \citet{2007ApJ...661..815R}, \citet{2008ApJ...683..611G}, \citet{2009AJ....138..332J}, \citet{2011A&A...531A.134T}, \citet{2011AJ....141..194G}, \citet{2012A&A...540A..49M}, \citet{2016AJ....151...88B}, \citet{2017AAS...23030103M}, \citet{2019MNRAS.490.5538A}, \citet{2020ApJ...899...67F}, \citet{2021ApJ...907..112L}, \citet{2021ApJ...916...19Z}, \citet{2023MNRAS.522..195P}.}
\label{fig:wlm-dist}
\end{figure}

\subsection{WLM} \label{sec:wlmdist}
We compared our result for WLM with the published distance results. A schematic compilation of the previous results post 2004 is presented in Figure \ref{fig:wlm-dist}. The published distance measurements of WLM exhibit good consistency overall and are consistent with our result as well. The mean value of the four Cepheid distances after 2007 is 24.94 mag, which is in good agreement with ours at the 1$\sigma$ level. The Cepheid distance determined in \citet{2007AJ....134..594P}, while relatively larger, remains consistent with our result within the 3$\sigma$ error range. The TRGB results agree well with our result except for \citet{2011AJ....141..194G} and \citet{2012A&A...540A..49M}. The total errors reported in these works are comparatively larger, which would probably account for the differences. All J-Region Asymptotic Giant Branch (JAGB) distances are in good agreement with this work within the 1$\sigma$ error range.

\begin{figure}[]
\centering
\includegraphics[width=\linewidth]{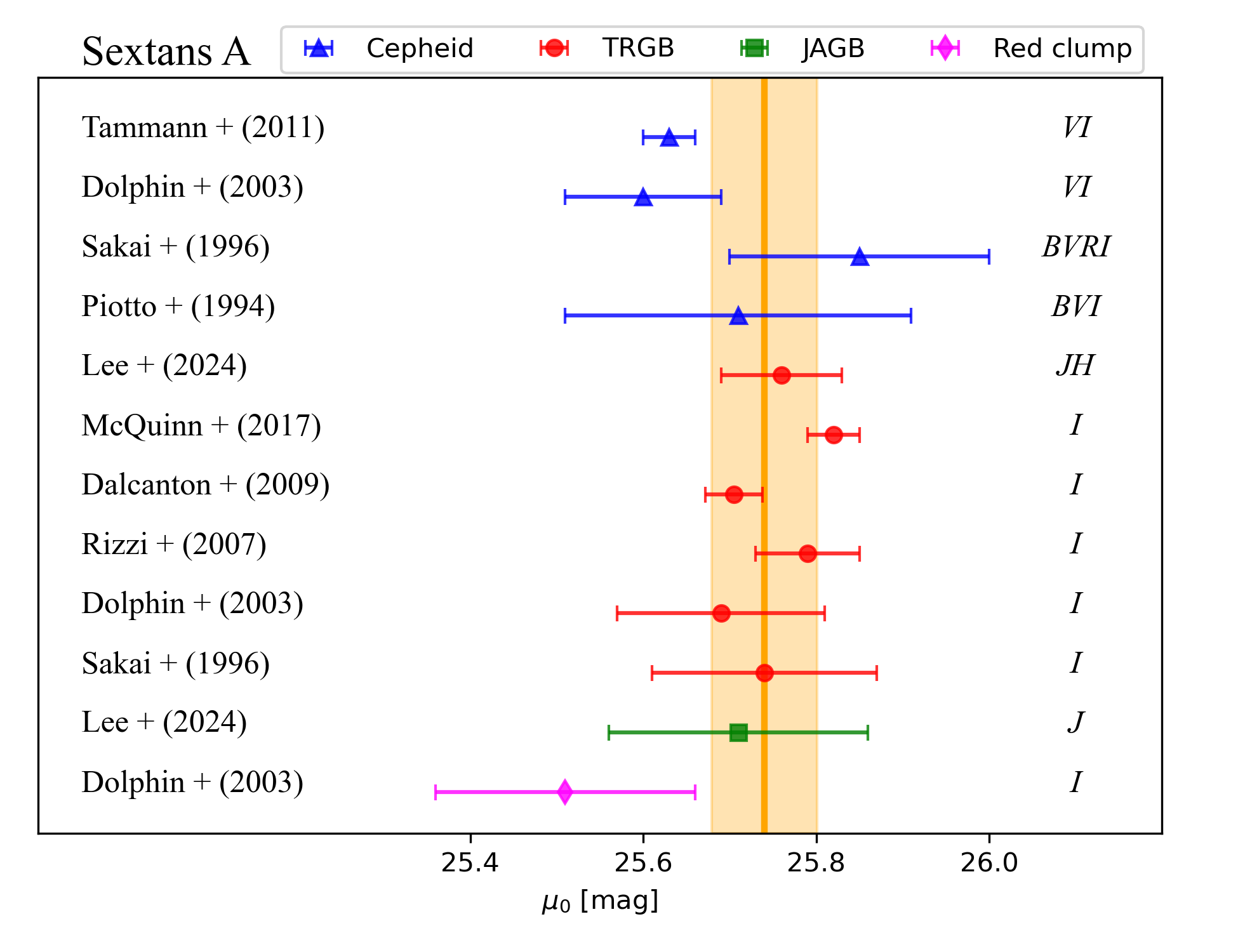}
\caption{Same as Figure \ref{fig:wlm-dist}, for Sextans A. References: \citet{1994A&A...287..371P}, \citet{1996ApJ...461..713S}, \citet{2003AJ....125.1261D}, \citet{2007ApJ...661..815R}, \citet{2009ApJS..183...67D}, \citet{2011A&A...531A.134T}, \citet{2017ApJ...834...78M}, \citet{2024ApJ...967...22L}.}
\label{fig:sexA-dist}
\end{figure}

\subsection{Sextans A} \label{sec:sexdist}
We made the same comparisons for Sextans A and a similar schematic compilation is shown in Figure \ref{fig:sexA-dist}. The published TRGB distances show excellent consistency, giving a mean distance modulus of 25.75 mag which agrees well with our result of 25.74 mag. The JAGB distance from \citet{2024ApJ...967...22L} also closely matches our TRGB result. The Cepheid distance from \citet{1994A&A...287..371P} agrees with our result at the 1$\sigma$ level. The Cepheid results from \citet{1996ApJ...461..713S} and \citet{2003AJ....125.1261D} are within the 2$\sigma$ error range of our result. The Cepheid result from \citet{2003AJ....125.1261D} is within the 3$\sigma$ error range of ours and shifts to 25.66 mag (1.3$\sigma$ level) after correcting the zero point of SMC to $\mu_{\mathrm{0,SMC}} = 18.93$ mag. The red clump modulus from \citet{2003AJ....125.1261D} is inconsistent with our TRGB result at the 3$\sigma$ level and smaller than all the other results, probably due to the greater influence of crowding effects to red clump stars.

Overall, the distances to WLM and Sextans A derived from multiple methods are consistent within 1$\sigma$ (corresponding to a $\sim$3\% uncertainty in distance). However, it should be noted that some previous measurements did not fully account for the complete sources of distance uncertainties.

\section{SUMMARY} \label{sec:summary}

In this work, we measured the distances to the dwarf galaxies WLM and Sextans A using the TRGB method based on JWST NIRCam observations. We constructed color–magnitude diagrams by applying spatial and color selection, and identified the TRGB locations in the F090W band using a Sobel edge-detection algorithm. The apparent TRGB magnitudes were measured to be $m_{\mathrm{F090W}} = 20.643 \pm 0.018$ mag for WLM and $m_{\mathrm{F090W}} = 21.415 \pm 0.011$ mag for Sextans A. Adopting the TRGB absolute magnitude $M_{\mathrm{TRGB}}^{\mathrm{F090W}} \approx -4.377$ mag calibrated from NGC 4258, we derived distance moduli of $\mu_{\mathrm{0,WLM}} = 24.977 \pm 0.018\ \mathrm{(stat)} \pm 0.056\ \mathrm{(sys)}$ mag and $\mu_{\mathrm{0,SexA}} = 25.740 \pm 0.011\ \mathrm{(stat)} \pm 0.057\ \mathrm{(sys)}$ mag, corresponding to distances of $0.989 \pm 0.027$ Mpc and $1.406 \pm 0.038$ Mpc, respectively, with an uncertainty of approximately 2.7\%. Compared to previous distance measurements based on HST TRGB, Cepheids, and JAGB methods, our results show good agreement within 1$\sigma$. Benefiting from the high signal-to-noise ratio and excellent spatial resolution of JWST, the distance uncertainties in this work are more comprehensively characterized than previous efforts.

Current TRGB-based $H_0$ measurements from different Type Ia supernova host galaxies show sample-dependent discrepancies. Therefore, detailed analyses of the TRGB magnitudes in nearby dwarf galaxies are also essential to assess the reliability of the existing TRGB calibration. In the future, new distance measurements to these dwarf galaxies may be achieved by RR Lyrae stars \citep{2023NatAs...7.1081C} based on more observations from JWST and China Space Station Telescope (CSST) and allow more differential comparisons of different distance indicators.

\section{Acknowledgements}
We thank the anonymous referee for the helpful comments. We thanked the support from the National Key Research and development Program of China, grants 2022YFF0503404. This research was supported by the National Natural Science Foundation of China (NSFC) through grants 12373028, 12173047, 12322306, 12233009 and 12133002. X. Chen and S. Wang acknowledge support from the Youth Innovation Promotion Association of the Chinese Academy of Sciences (CAS, No. 2022055 and 2023065). This work is based on observations made with the NASA/ESA/CSA James Webb Space Telescope. The data were obtained from the Mikulski Archive for Space Telescopes (MAST) at the Space Telescope Science Institute.

\bibliography{reference}{}
\bibliographystyle{aasjournal}

\end{document}